# Platform for generating medical datasets for machine learning in public health

Anna Andreychenko, Viktoriia Korzhuk,
Stanislav Kondratenko, Polina Cheraneva

aeandreychenko@itmo.ru, vmkorzhuk@itmo.ru,
stanislav.kondratenko98@gmail.com, apolloisherebynow@gmail.com

«Digital Public Health Technologies» Laboratory
ITMO University
Saint-Petersburg, Russian Federation

**Abstract**

Currently, there are many difficulties regarding the interoperability of medical data and related population data sources. These complications get in the way of the generation of high-quality data sets at city, region and national levels. Moreover, the collection of datasets within large medical centers is feasible due to own IT departments whereas the collection of raw medical data from multiple organizations is a more complicated process. In these circumstances, the most appropriate option is to develop digital products based on microservice architecture. Because of this approach, it is possible to ensure the multimodality of the system, the flexibility of the interface and the internal system approach, when interconnected elements behave as a whole, demonstrating behavior different from the behavior when working independently. These conditions allow, in turn, to ensure the maximum number and representativeness of the resulting data sets. This paper demonstrates a concept of the platform for a sustainable generation of quality and reliable sets of multimodal medical data. It collects data from different external sources, harmonizes it using a special service, anonymizes harmonized data, and labels processed data. The proposed system aims to be a promising solution to the improvement of medical data quality for machine learning.

## 1 Introduction

For the past few years, healthcare has experienced significant growth in clinical capabilities and access to care by actively involving machine learning (ML) [1]. However, valuable ML applications in clinical practice are still incredibly limited. The number of ML systems that are being developed and then deployed in clinical settings is increasing. There are a large number of projects that have successfully trained machine learning models and are getting good results in experimental studies. However, one of the main difficulties of such developments, that the world scientific community considers a priority in this area, is insufficient attention to the interface and, accordingly, the complexity of directly using of trained machine learning models in practical applications [1,2]. Most of the models don't stand the test of real-world data and that is the reason for the paradigm shift to data-driven ML.

There are a lot of approaches and examples of centralized biomedical data collection systems that are used all around the world, however, these solutions are not universal and do not have a microservice or modular architecture as a result it restrains their scalability and flexibility. Moreover, most of these systems were created for biomedical studies without considering machine learning requirements [2].

One of these requirements touches on the problem of data preparation which also implies the task of data collection. Most of the time doctors are occupied with patients so they do not have time to collect data (unless it is a clinical trial), and there are no formal or documented rules on how to do that. Additionally, the data should be depersonalized or kept in secure storage and transmitted using a secure communication channel [3]. It is necessary to obtain patient consent to the processing of personal data [3]. Besides, every doctor involved in the data collection and data labelling should be qualified enough to do so [4]. The standards indicate that doctors involved in marking up medical data should be competent in the field of specific types of data: images, text data or signal data (for example, ECG, EEG), quantitative data (heart rate, blood pressure), binary data (yes or no). Also some addition factors have to be taken into account: the level of complexity of the planned labelling or annotation (primary labelling (segmentation) or expert one), as well as the level of detailing, determining probable outcomes (prediction). Finally, doctors have to pass successfully some specific tests. The expert group that controls the quality of the labelling also must meet the requirements: extensive experience working with a certain type of thematic datasets (usually more than 3 years)..

Some requirements also revolve around the data itself. Any data given to the interns or residents can be used for manual processing. But this approach does not work from the ML perspective and we are developing a solution to this problem. To be specific, the format of the input data should be unified and easy to understand, as well as protocols, text fields, characteristics and others should be flexible, scalable and at the same time standardized for each clinical task.

Furthermore, during an ML model development outliers are commonly excluded from datasets, especially from the training samples. But in medicine, outliers are important since they may serve as indicators of rare diseases particularly significant for public health and medicine development [5,6].

Hence, our goal was to develop and then implement a microservice architecture of the system that stores, processes and labels multimodal medical data as well as satisfies the requirements for datasets enforced by ML.

The paper is structured into five sections. Firstly, we highlight the problem of centralized biomedical data systems, briefly overview current globally used solutions and introduce our approach to this problem. The next section presents a detailed overview of different open-source solutions in medicine. The architecture section describes an overall structure of the system and its components. After we discuss impacts of the system on AI applications in medical practice and other benefits of our project. Finally, the conclusion section explores prospects and implications of the platform.

## 2 Overview of open source solutions

The development of various open-source solutions in medicine can significantly impact the evolution of medical systems. Moreover, the use of open-source software helps to disseminate knowledge and increase the level of competence in the scientific community [7]. The purpose of the overview is to select the most suitable instruments and methods to work with medical data.

At the moment 120 open-source solutions were analyzed. They were divided into 2 categories according to data types and stages of life cycle.

### 2.1 Life cycle stages

Four stages of the life cycle were identified: data collection, data labeling, development of ML models, and data storage.

12 different data collection tools related to medical data were reviewed, for example TorchXRayVision[13]. As well as 14 data labeling sources were analyzed including CVAT[14] and Slicer[15]. 22 different sources for AI development, depersonalization, and database creation for storage and processing of medical data were observed, in particular, MedicalNet[17] and HIMA[16]. After analyzing, it was clarified that 35 of the methods were implemented using Python, 10 of them are cross-platform and support Java or C++, and 15 of them have not been updated since August 2022.

The most popular universal solutions for data processing for different stages of a life cycle are clearML, MLflow, Optuna, Tuva Project, etc [8,9,10,11]. However, due to the specifics of the field these tools need to be adapted to successfully solve medical tasks. For example, some of analized tools have poor UI/UX design, some require a lot of development experience for installation and deployment, which makes it cumbersome to use in practice in medical institutions; some solutions support only certain browsers and labelling tools, some have limitations on the format of processed and stored data, and others work only offline. Some tools work only with previously depersonalized data, and some were developed for a specific clinical task and cannot be expanded.

Several well-known repositories storing datasets were also analyzed - the well-known Mosmed-AI resource [12] was among them. Analysis of the data sets stored there allowed us to get a real impression of a high-quality data set that can be used to solve applied problems.

The analysis revealed 4 categories of medical data that were used in the reviewed software: laboratory data, electronic health records (text and table data), medical images, and ancillary data. Typical formats of medical images were DICOM, TIFF, BMP, and NIFTI. As for text and table data, common formats were XML, CSV, JSON, and TXT. Figure 1 illustrates the tools distribution according to life cycle stages and data types.

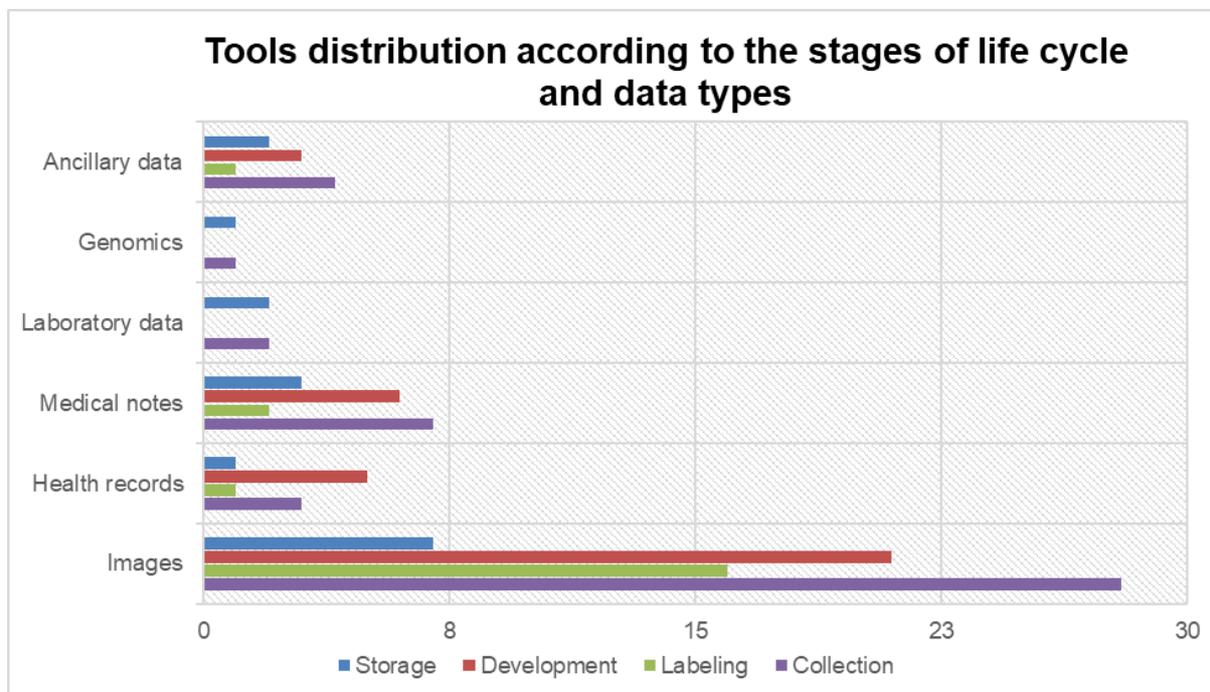

Figure 1- Diagram of tools distribution according to life cycle stages and data types

**2.2 Data types**

Types of open-source solutions can be categorized by their application areas such as image processing, processing of natural language, comprehensive solutions, and completed models (assembled models ready for training and application in clinical tasks).

The graph in figure 2 shows the distribution of application of open-source solutions according to different fields. 57% of the sources are used to process images, 17% to process natural language, 25% in comprehensive solutions, and 10% in completed models.

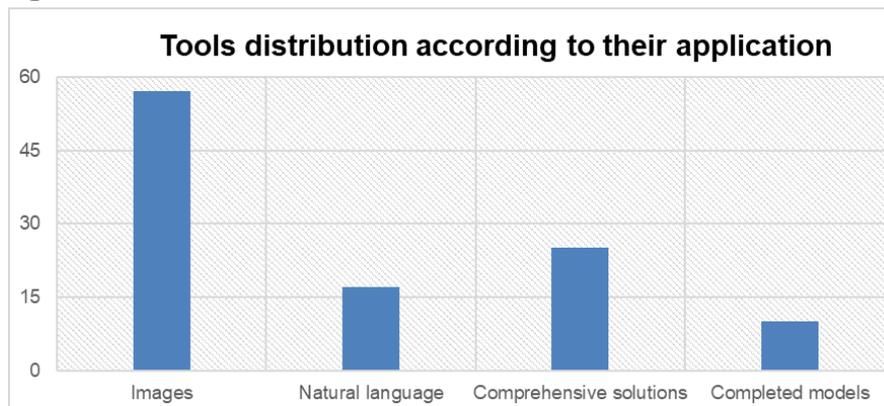

Figure 2 - Diagram of tools distribution

As for the most common image formats, the pie chart in figure 3 illustrates that 40 % of the total number of images are stored in the DICOM format. Second most frequent formats are PNG and JPEG, their share is 15%. NIFTI is used in 11 % of cases, while TIFF is only used in 5% of situations. PDF format is extremely rare, its share is 1% .

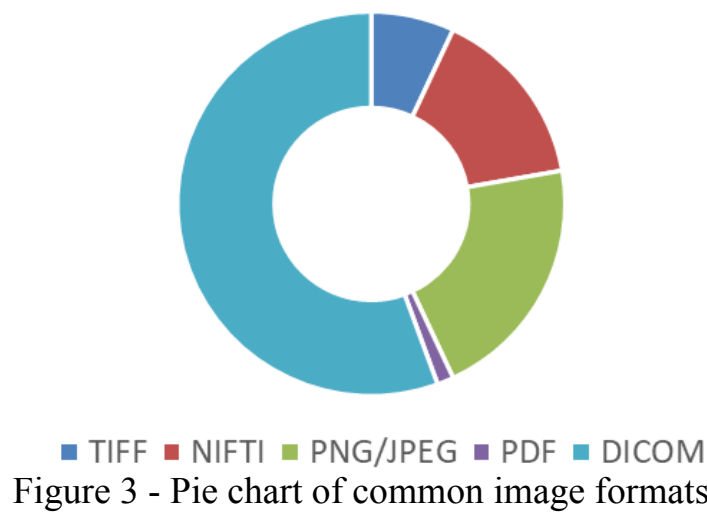

Figure 3 - Pie chart of common image formats

# 3 Proposed Platform

The main goal of the platform is to collect high-quality multimodal data sets based on the main tasks of doctors without a large amount of additional load. At the current stage of the project development, we are forming a data set from medical images and collecting text data and other data formats separately, but we plan to link these processes and form multimodal data sets in the true meaning.

The proposed platform integrates into the usual data labeling procedures and allows doctors to save and view data. This system integrates into the usual procedures of data labeling and allows doctors to save and review the data. At the moment, the system is intended to work with data of a lung cancer screening programme. We implemented our data labeling tool and depersonalization algorithm which is currently undergoing collision and security tests. Moreover, the proposed system provides many different ways to store data. It can be stored in a cloud, in centralized data storage, or distributed in local storages of the organizations that developed a specific database for screening.

This section overviews main components of the system such as data sources, microservices, queues, and data warehouse, and discusses the data flow on the example of the lung cancer screening task. Figure 4 presents the whole scheme of the proposed system.

## 3.1 Platform Structure
**Sources**

Several different external sources transmit data to the system such as CRM systems (b24 register) that provide a register of members, protocols of the radiology information system (RIS) that forward data of radiological tests and patients diagnosis, Electronic health records (EHR) that keep overall health information about patients, and external picture archiving and communication system (PACS) that stores DICOM images. Data from CRM, EHR, and RIS protocols is sent to the system microservices whereas the data from the external PACS is transmitted to the PACS router.

**Microservices and queues**

Each source is assigned an individual microservice that grabs its data using a specific script and distributes it to the corresponding topic in the queue.

Queues allow the system to provide security, consistency, and integrity of data. For instance, in case of an attack that can put the system out of action all the transmitting data at the moment of shutdown will be stored in the queues and after reboot the system will be able to continue work where it was left off. In this project, queues are implemented by using Apache Kafka.

The first topic of the first queue receives data from the register of members from the CRM system, data on radiological tests from RIS protocols is sent to the second topic, and general health information of patients from the EHR is distributed to the third topic.

**Data warehouse**

The data warehouse consists of several parts such as data aggregation microservice, internal PACS, depersonalized register, data labeling tool, and business intelligence (BI). The data labelling tool has access to all received images (DICOMs) that are stored in the internal PACS. This tool provides the opportunity to label images and fill in medical protocols according to the labeling results. Once the filling of the protocol is completed it is automatically saved in the depersonalized register.

Using business intelligence tools, a doctor can study detailed information about each patient or study general statistics for all patients. The BI tool is a layer between the doctor and the data labeling tool that allows you to flexibly customize the displayed data, display a graphical interpretation of dependencies and mark important elements.

By a patient or a screening participant, we mean a person who independently applied to a medical organization and provided data on their health status. As part of the development of the platform, a special form was also prepared for the registration of screening participants, which allows automating the process of verifying the patient's compliance with the screening conditions. Each patient after the examination is assigned with one out of three possible results: no signs of malignant neoplasms of the lungs, needs medical supervision, and needs additional examination. The examination is carried out according to the Lung-RADS standards. If the result is «no signs of malignant neoplasms of the lungs» then the participant is marked healthy in the system, if the patient needs medical supervision then this person will be invited to the screening again after a set period of time, and if the participant needs additional examination then he or she is immediately informed via call.

Based on how complex the particular case is, doctors involved in the labeling have an opportunity to get a second expert opinion by inviting an external medical specialist. In the depersonalized register the participant is marked as needing a second medical review. The invited specialist can open the image, re-label it and fill in the protocol.

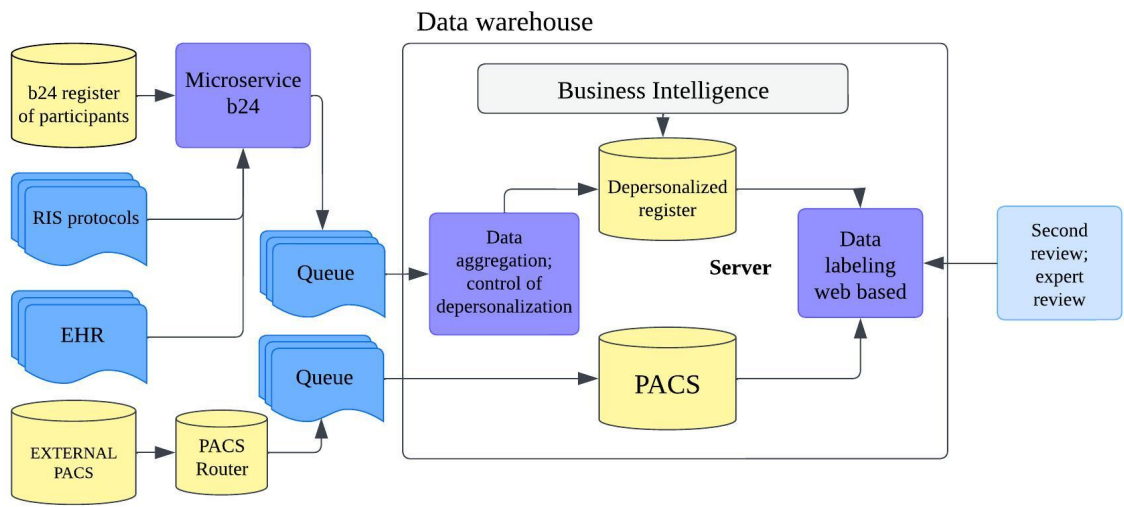

Figure 4 - Platform structure

## 3.2 Data flow

Figure 5 demonstrates the context diagram of the system that depicts and explains relations, dependencies, and interactions between entities and their environment (people, systems, and external sources). This diagram is presented as a simplified highest-level picture showing the system boundaries and interfering external objects.

The system is separated from data sources, doctors, and screening organizers. DICOM images and text information in TXT and CSV formats are transmitted from external sources by using a data harmonization service that consists of the PACS router and system microservices. Microservices grab data in text format by executing special triggers, the PACS router gathers DICOM images from external PACS, and after harmonization, the data is sent to the data storage.

After the data arrives at the storage it is available for service for orchestration of medical test interpretations and web application for interpretation of medical tests. The orchestration service provides the opportunity for the doctor to examine information about a patient in chronological order starting from general information, and medical diagnoses, and ending with medical examination protocols. As a web application for medical image viewer, it is proposed to use OHIV Viewer for examining and labelling DICOMs. A special add-in allows to automatically generate protocols of the established form on the conducted examination.

At the end processed and labeled data in the tabular form is transmitted from the storage to the screening information panel for a data analysis.

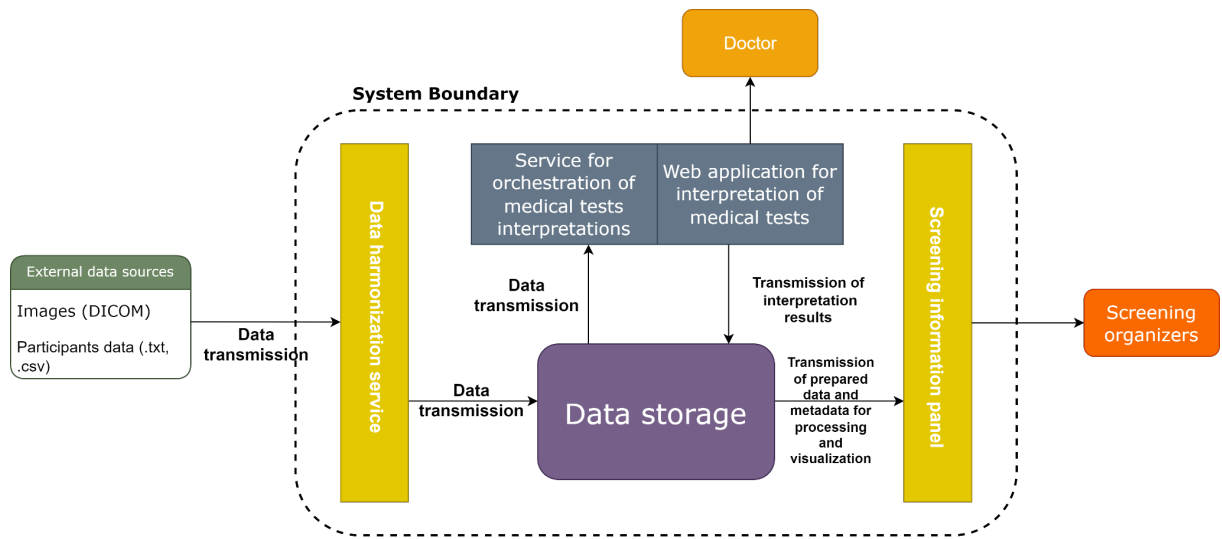

Figure 5 - Context diagram of the system

## 4 Discussion

In the future, the system will enhance medical data quality for machine learning algorithms. The developed prototype is a necessary step to ensure continuous and reliable formation of datasets. Such a platform combined several sources of medical data and provided the ability to store essential data in one place. This allows to streamline the screening program, scale it regardless of the number of medical institutions.

Flexibility and the possibility of further improvement and collaboration on the platform with other teams of researchers and developers, for example, engaged in the use of artificial intelligence in the analysis of medical data, make this platform universal and relevant. AI application in medicine can improve the efficiency of medical care and reduce doctors' workload. Routine process automation and big data analysis allow doctors to focus on complex and critical problems increasing the quality of treatment.

Other important aspects are the generation of reference datasets and the development of the platforms that will be able to process them. These developments will help with the standardization process and upgrade sharing of medical data by increasing the accuracy of diagnosis and decision-making.

On top of that, the development of a universal solution with microservice architecture provides the opportunity to ensure flexibility and scalability of the AI system in medicine. It contributes to the effectiveness of integration and collaboration of various modules and applications by sustaining efficient use of AI for different tasks in medical practice.

Naturally, the current solution has limitations. For example, when scaling the architecture, there may be issues of data processing speed and anonymization, but it can be solved on the basis of carefully configured orchestration. The issue of forming a unique identification number of a patient and his research inside the platform and his connection with an account inside a medical institution has not been fully investigated.

The platform can be easily adapted to various types of screening studies and has the ability to easily connect intelligent data analysis tools. Potentially, the platform can be used as an analogue of a medical information system and contribute to the personalization of medical care.

Speaking globally, the platform can potentially become part of the state medical information system or become a theoretical basis for the development of projects in the field of the formation of reference sets of medical data.

## 5 Conclusion

In this work, we presented a platform prototype for generating medical datasets for machine learning in public health. The project is implemented using open source technologies. The microservice architecture allows us to flexibly configure and scale our system for the tasks of a specific medical institution. The implemented system allows to organize: secure transfer of medical data, work with images with a large set of tools in the web labelling, as well as the formation of the final protocol of the examination for the patient, screening participant, the doctor and the expert. Proposed platform can be deployed by a doctor independently for his personal purposes, as well as implemented for the needs of any medical institution for specific tasks. At the moment, the configuration of the following components has been completed: PACS (orthanc), web markup and research protocol (ohif), database (PostgreSQL). All components are interconnected and meet the basic requirements of information security - confidentiality, integrity and accessibility.

The source code for the platform can be found in the following GitHub repositories: https://github.com/AlpHaDPHT-SE-Department/ - this represents the whole project on proposed platform by youth scientific laboratory «Digital Technologies in Public Health» (https://dpht.itmo.ru) from ITMO University, and https://github.com/AlekseiHead/Viewers/tree/modal-form-with-formik represents our team results in web-viewer development.

The next stage is the deployments of the product on the side of a partner medical organization. It is planned to assess the reduction of labor costs by automating a number of processes related to the routing of screening participants and doctors, improving the effectiveness of medical care due to more timely and prompt diagnostics and prognostication. It is also planned to evaluate the impact of the platform on the quality of medical data for machine learning algorithms by increasing the amount of high-quality markup on real cases.

**Acknowledgements**

This study was supported by grant FSER-2022-0013 of the national project "Science and Universities."